\documentclass{INTERSPEECH2023}

\interspeechcameraready

\usepackage{multirow}
\usepackage[bottom]{footmisc}

\title{What Can an Accent Identifier Learn? Probing Phonetic and Prosodic Information in a Wav2vec2-based Accent Identification Model}
\name{Mu Yang$^1$, Ram C. M. C. Shekar$^1$, Okim Kang$^2$, John H. L. Hansen$^1$}
\address{
  $^1$Center for Robust Speech Systems (CRSS), University of Texas at Dallas, Richardson, TX, USA\\
  $^2$Northern Arizona University, Flagstaff, AZ, USA}
\email{$^1$\{mu.yang, ramcharan.chandrashekar, john.hansen\}@utdallas.edu,  $^2$okim.kang@nau.edu}

\begin{document}

\maketitle
 
\begin{abstract}
% 1000 characters. ASCII characters only. No citations.
This study is focused on understanding and quantifying the change in phoneme and prosody information encoded in the Self-Supervised Learning (SSL) model, brought by an accent identification (AID) fine-tuning task. This problem is addressed based on model probing. Specifically, we conduct a systematic layer-wise analysis of the representations of the Transformer layers on a phoneme correlation task, and a novel word-level prosody prediction task. We compare the probing performance of the pre-trained and fine-tuned SSL models. Results show that the AID fine-tuning task steers the top 2 layers to learn richer phoneme and prosody representation. These changes share some similarities with the effects of fine-tuning with an Automatic Speech Recognition task. In addition, we observe strong accent-specific phoneme representations in layer 9. To sum up, this study provides insights into the understanding of SSL features and their interactions with fine-tuning tasks.
\end{abstract}

\section{Introduction}

Second-language (L2) English learners typically present their first-language (L1) accents (e.g. Mandarin-accented English). An accent identification (AID) model seeks to identify the L1 accent of a speaker given his/her spoken L2 utterance. Modeling accents in speech has many practical applications. Knowledge of accents has been shown to benefit Automatic Speech Recognition (ASR) applications \cite{weninger2019deep, gao2021end, shi2021accented, biadsy2011automatic} and Computer-aided Pronunciation Training (CAPT) scenarios \cite{witt2012automatic, saito2022automated}, enabling more robust and inclusive speech technologies.  

In general, accent refers to a particular pronunciation style of a spoken language, which can be characterized by both segmental and suprasegmental features \cite{witt2012automatic}. On the segmental level, accent can be perceived by certain phonological errors, such as mispronounced phonemes and articulation variations \cite{yang22v_interspeech, wu2021transformer, suzukida2021segmental}. On the suprasegmental level, accents are usually associated with inaccuracies in prosody factors, such as lexical stress, intonation, and rhythm \cite{kang2010suprasegmental, ulbrich2016prosody, li2018automatic}. 

Recently, there have been extensive studies on end-to-end modeling of AID \cite{gao2021end, deng21b_interspeech, wang2021end, shon18_odyssey}. However, relatively limited research has been done to \textit{understand} and \textit{quantify} the segmental and suprasegmental information encoded in the AID model. The present study fills this gap by analyzing the intermediate representations of an AID model on both phonetic and prosodic aspects. 
Specifically, we build an AID model by fine-tuning the base version of Wav2vec 2.0 \cite{baevski2020wav2vec}, a Self-supervised Learning (SSL) model that is pre-trained on a large volume of unlabeled audio. Recent works have shown that SSL models encode phonetic \cite{pasad2021layerwise,dieck22_interspeech,ma2021probing,cormac-english-etal-2022-domain}, prosodic \cite{lin2022on} and speaker-related information \cite{feng2022silence, fan21_interspeech}.
The primary research question we ask is: how would AID, an \textit{utterance-level} fine-tuning task guide the segmental-level (i.e. phoneme) and suprasegmental-level (i.e. prosody) representation change in the SSL model? 
To approach this problem, we design two probing tasks (one phoneme-related and one prosody-related), and conduct layer-wise analyses on the representations of the intermediate Transformer \cite{vaswani2017attention} layers. This method enables us to 1) inspect how the phonetic and prosodic information is encoded and distributed in the AID model; 2) quantify the change of the encoded phonetic and prosodic information brought by the AID fine-tuning task, by comparing the performance of the probing tasks between the pre-trained Wav2vec 2.0 and the fine-tuned AID model.

In this study, we consider two probing tasks. For phoneme probing, we use Canonical Correlation Analysis (CCA) \cite{pasad2021layerwise} to measure the similarities between the hidden representations and phoneme targets. For prosody probing, we propose a novel regression-based task to predict the word-level prosodic prominence and boundary scores \cite{suni2017hierarchical}. Prominence and boundary are perceived due to the local increase or reduction of one or more of the prosodic attributes such as duration, F0, energy, and spectral shape. Hence, they are reasonable measurements of prosodic structures. We probe 3 models: 1) the pre-trained Wav2vec 2.0; 2) the Wav2vec 2.0 model fine-tuned by an AID task; 3) the Wav2vec 2.0 model fine-tuned by an ASR task. We hope that this study can provide insights into the understanding of SSL features
and their interactions with fine-tuning tasks. Our primary findings are summarized as follows:
\begin{itemize}
    \item AID fine-tuning mainly causes changes in the top 4 layers. 
    \item The top 2 layers gained improved representations for both phoneme and prosody, and these changes share some similarities with the effects of ASR fine-tuning.
    \item In layer 9 we observe strong accent-specific phoneme representations.
\end{itemize}

\section{Related work}
In this section, we review prior works that analyze the intermediate representations of speech models.
Belinkov and Glass \cite{belinkov2017analyzing} designed classier-based probes to analyze the hidden representations of an end-to-end ASR model, Deep Speech 2 \cite{amodei2016deep}. Prasad and Jyothi \cite{prasad2020accents} further investigated how accented speech segments affect the hidden representations and outputs of Deep Speech 2. 

More recently, analyses of SSL models have gained increased attention. On the phoneme side, prior works inspected Wav2vec 2.0 models fine-tuned by phoneme recognition \cite{dieck22_interspeech} or ASR tasks \cite{pasad2021layerwise}. Our work differs from theirs in that we investigate how an utterance-level AID task steers the representation change in the SSL models. Similar to our work, Triantafyllopoulos et al. \cite{triantafyllopoulos22b_interspeech} fine-tunes Wav2vec 2.0 with an utterance-level emotion recognition task, and probe the linguistic knowledge in the fine-tuned model, while we focus on a different set of probing tasks. On the prosody side, a recent study \cite{lin2022on} evaluates SSL models' downstream performance on prosody-intensive tasks such as sentiment analysis and sarcasm detection, as well as prosody feature (pitch and energy) prediction. In contrast, we directly probe the SSL representations in a layer-wise manner. Moreover, instead of predicting the lower-level pitch and energy, we focus on higher-level prosodic prominence and boundary features, because they are associated with linguistically relevant units (e.g. phonemes, words). Such features are more related to overall speech fluency \cite{vaidya2022deep} and serve as better indicators in the context of L2 speech.

\section{Materials and Method}

\subsection{Datasets}
We merge ARCTIC \cite{kominek2004cmu} and L2-ARCTIC \cite{zhao2018l2} datasets for AID training and probing. L2-ARCTIC includes L2 English speech from 24 non-native English speakers with 6 L1 backgrounds: Indian (IN), Mandarin (CN), Vietnamese (VN), Korean (KR), Arabic (AB), and Spanish (SP), with 2 male and 2 female speakers per accent. We also include utterances from the 4 US-accented speakers (CLB, RMS, BDL, SLT) in ARCTIC, leading to 7 accents, 28 speakers in total. For AID training, we held out 7 speakers (one for each accent) as the test set: SKA, BWC, SVBI, HKK, NJS, HQTV, SLT. Data statistics for AID training are shown in Table \ref{tab:data_stats}. 
% Probing experiments are done on the same datasets. 
More data usage details about the probing experiments are discussed in Section \ref{sec:probe_phone} and \ref{sec:probe_prosody}.

\begin{table}[t]
    \caption{Data statistics for AID fine-tuning and probing tasks.}
    \label{tab:data_stats}
    % \vspace{-3mm}
    \centering
    \resizebox{1.0\columnwidth}{!}{
    \begin{tabular}{c|clcl}
\hline
\textbf{Experiments}                      & \multicolumn{2}{c|}{\textbf{Train}}                & \multicolumn{2}{c}{\textbf{Test}}                \\ \hline
\multirow{3}{*}{AID Fine-tuning} & \multicolumn{2}{c|}{\# speakers: \SI{21}}      & \multicolumn{2}{c}{\# speakers: \SI{7}}      \\
                                 & \multicolumn{2}{c|}{\# utterances: \SI{23633}} & \multicolumn{2}{c}{\# utterances: \SI{7762}} \\
                                 & \multicolumn{2}{c|}{\# hours: \SI{60.3}}       & \multicolumn{2}{c}{\# hours: \SI{18.3}}      \\ \hline
Phoneme Probing                  & \multicolumn{4}{c}{$\sim$\SI{70}{k} segments (3-fold Cross Validation)}                                  \\
Prosody Probing                  & \multicolumn{4}{c}{$\sim$\SI{278}{k} segments (4-fold Cross Validation)}                                 \\ \hline
\end{tabular}
}
\vspace{-3mm}
\end{table}

\subsection{The accent identification model}
\label{sec:aid_finetune}
We build a simple AID model by fine-tuning Wav2vec 2.0. Specifically, we fine-tune the pre-trained \textit{wav2vec2-base} checkpoint on an utterance-level accent classification task. The output of Wav2vec 2.0 is sequentially processed through layer normalization, a 256-dim single linear projection layer, and a dropout module with a dropout rate 0.1. We then use mean pooling along the time dimension to get utterance-level representation, which is then passed to the final linear projection layer to produce the logits for the 7 accent classes. We use Cross Entropy with a label smoothing factor \cite{szegedy2016rethinking} of 0.25 as the loss function. Label smoothing softens the hard one-hot targets, and we observe that label smoothing gives a more balanced classification accuracy across all accents. 
It was also shown helpful to utterance-level emotion recognition \cite{kakouros2022speech}.

Our initial experiments show that although we use accent labels as fine-tuning targets, the AID model actually strongly relies on speaker identity to make accent predictions. This leads to high training accuracy, but very poor generalization when identifying accents for unseen speakers. To break the dependency on speaker identity, we apply on-the-fly perturbations on the input waveform to alter speaker voices \cite{choi2021neural, qian2022contentvec}. Specifically, the perturbations consist of three sequential random transformations: 1) all the formant frequencies within an utterance are scaled by a factor of $\beta_1$; 2) F0 is scaled by a factor of $\beta_2$; 3) a random frequency-shaping equalizer. $\beta_1$ and $\beta_2$ are both randomly drawn from the uniform distribution $\mathcal{U}(1, 1.4)$, and then flipped to their reciprocals with probability 0.5. Since the scaling and frequency shaping are all uniformly applied to the utterance, such perturbations mainly alter the speaker's voice while minimally affecting the spoken content or accent. During fine-tuning, for each batch, we draw a random variable $\alpha$ from a uniform distribution $\mathcal{U}(0, 1)$. The perturbations are applied if $\alpha > 0.25$. Otherwise, the original waveforms are used instead.

\subsection{Phoneme probing: Canonical Correlation Analysis}
\label{sec:probe_phone}
Canonical Correlation Analysis (CCA) measures the maximum correlation between the linear projections of two continuous-valued vectors. Pasad et al. \cite{pasad2021layerwise} proposed to use CCA as a similarity measure between hidden representations and phoneme labels. Specifically, given a set of vector pairs $\{(x_1, y_1), (x_2, y_2), \ldots, (x_n, y_n)\}$, where $x_i \in \mathbb{R}^{d_1}$ denotes the hidden representation for a phoneme,  $y_i \in \mathbb{R}^{d_2}$ represents the corresponding one-hot encoded phoneme label, the \textit{directions} of maximum correlation between linear projections of $X$ and $Y$ are defined as $v_1, w_1=\arg \max _{v, w} \operatorname{corr}\left(v^T X, w^T Y\right)$. The subsequent directions $v_i, w_i, \forall i \in[2, \min (d_1, d_2)]$ are required to be uncorrelated while maximizing the same correlation objective. We employ the same projection-weighted CCA variant in \cite{pasad2021layerwise} that computes a weighted sum of $\rho_i$s, where $\rho_i = \operatorname{corr}\left(v_i^T X, w_i^T Y\right)$. To extract phoneme representations from the SSL model, we use Montreal Forced Aligner \cite{mcauliffe17_interspeech} to obtain phoneme time stamps, and average the hidden representations within the central third of the intervals. We randomly sample 100 segments per phoneme in the dictionary, for each speaker across the entire ARCTIC plus L2-ARCTIC corpus. We discard segments that are shorter than 2 frames. This leads to $\sim$\SI{70}{k} phoneme segments. The final CCA scores are computed in a Cross Validation-like style: the probing samples are split into 10 folds, one for testing and the rest for training $v_i$s and $w_i$s. The average of CCA scores on 3 test folds is reported.

\subsection{Prosody probing: prominence \& boundary prediction}
\label{sec:probe_prosody}
We also aim to investigate the prosody knowledge encoded in the hidden representations of the AID model. We choose to use word-level prosodic prominence and boundary as our target features. These two features primarily quantify the chunking of speech into linguistically relevant units and the relative salience of the given units. They are computed based on the pitch, energy, and duration properties of the units. A local increase in the properties reflects a prominent unit, while reductions in the properties may indicate boundary units \cite{suni2017hierarchical}. We use an open-source tool\footnote{\url{https://github.com/asuni/wavelet_prosody_toolkit}} to label the prominence and boundary scores for all the words in the entire ARCTIC plus L2-ARCTIC datasets (two continuous scalar scores for each word, one for prominence, one for boundary). We then add a regression head (single linear layer of output dimension 1) on the word-level representations to fit the target values using Mean Squared Error (MSE). Word-level representations are extracted by averaging the representations within the word intervals provided by forced alignment.\footnote{We also tried phoneme-level prominence and boundary prediction. The results follow similar trends as word level.}

We split the sampled words into 4 folds, where each fold has 7 speakers of 7 accents. We conduct Cross Validation on the 4 folds, in which we test on 1 fold and train on the rest 3 folds 4 times. Average MSE of 4 test folds is reported.

\begin{figure}[b]
    \centering
    \vspace{-5mm}
    \includegraphics[width=1.0\columnwidth]{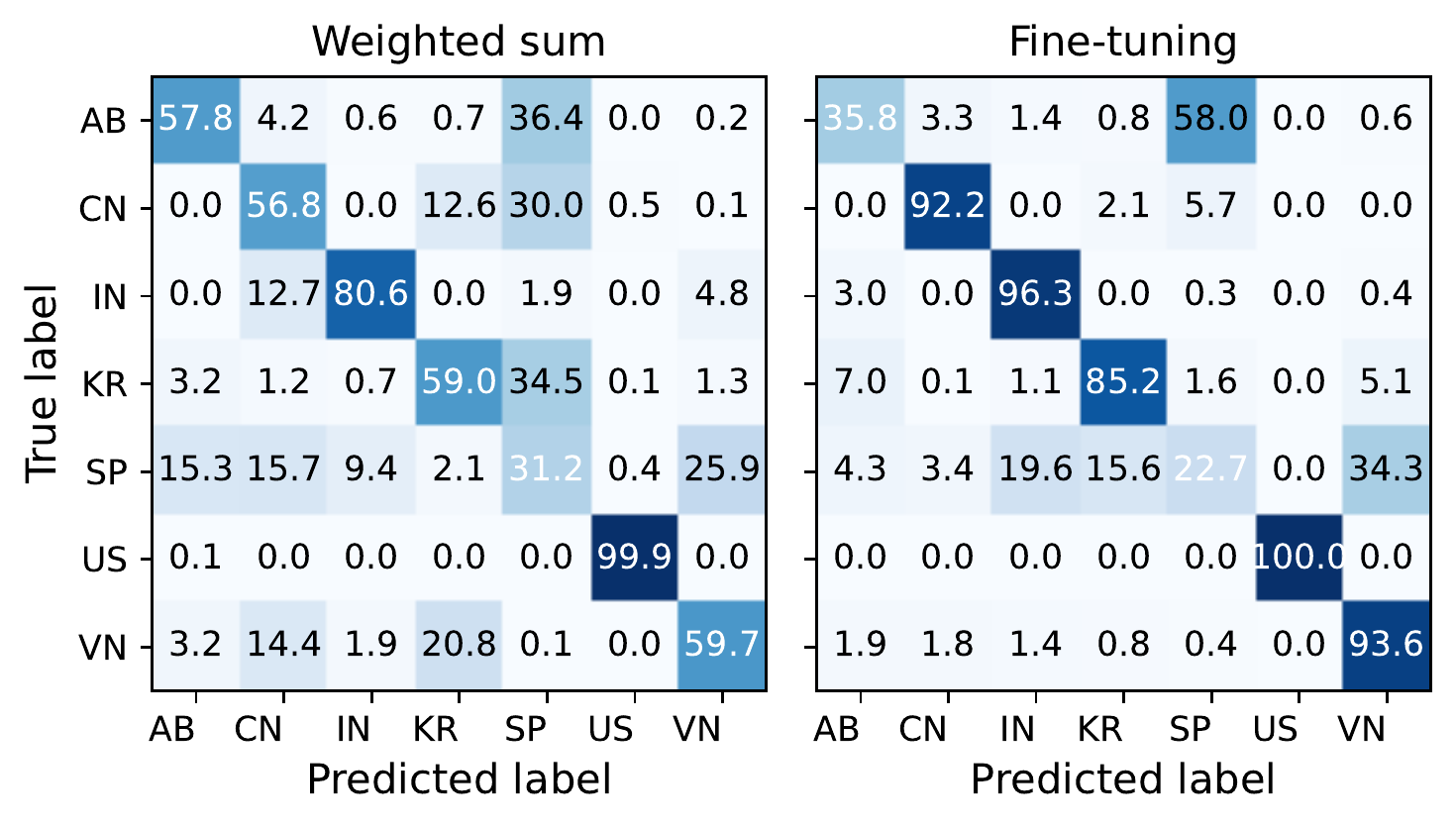}
    \vspace{-5mm}
    \caption{Accent identification accuracy (\%). The overall average accuracy for weighted sum and fine-tuning is 63.7\% and 75.9\%, respectively.}
    % \vspace{-5mm}
    \label{fig:aid}
\end{figure}
\begin{figure}[t]
    \centering
    \includegraphics[width=1.0\columnwidth]{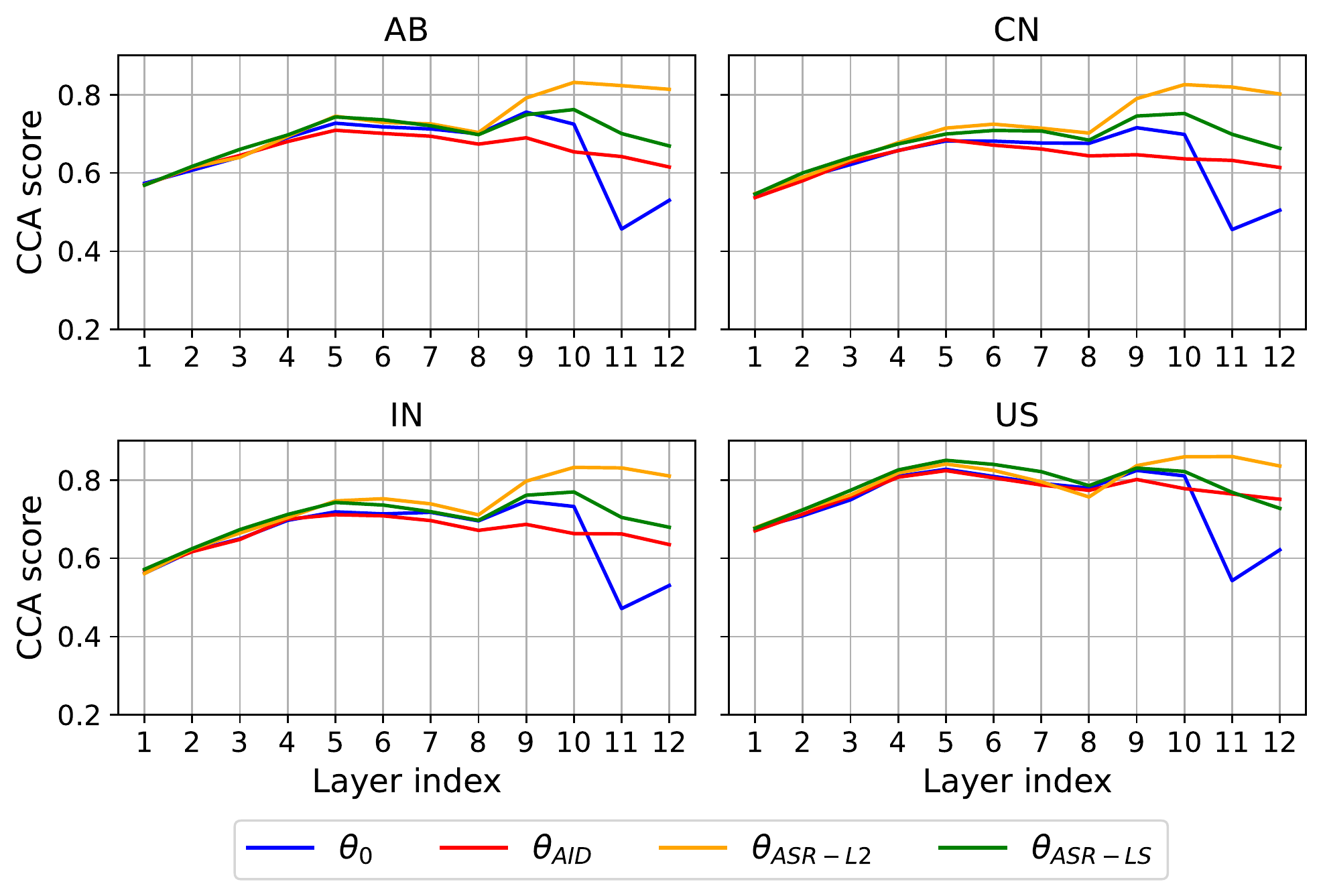}
    \vspace{-5mm}
    \caption{Layer-wise CCA on AB, CN, IN, and US accents.}
    \vspace{-5mm}
    \label{fig:cca}
\end{figure}
\begin{figure*}[t]
    \centering
    \includegraphics[width=1.0\textwidth]{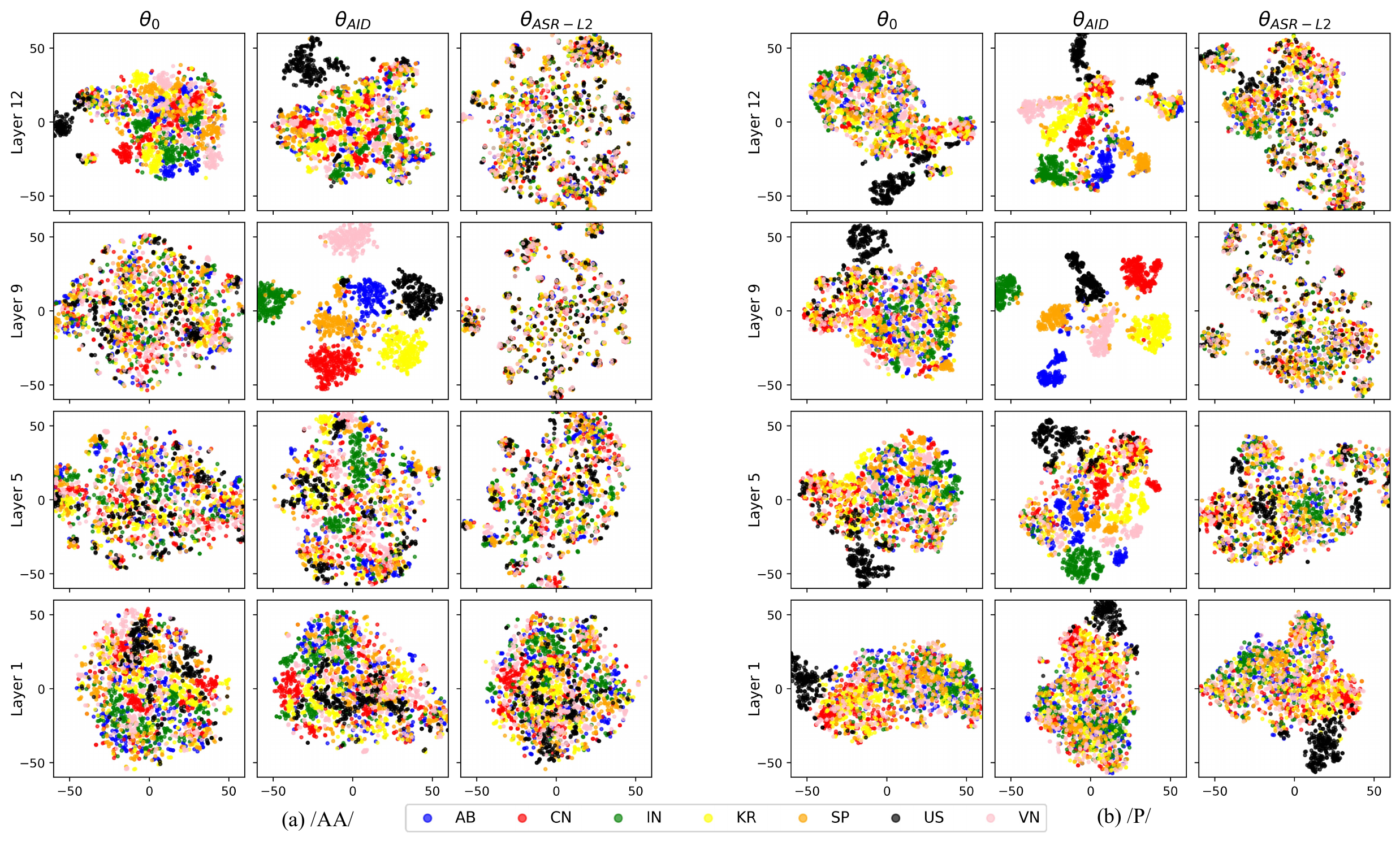}
    % \vspace{-6mm}
    \caption{t-SNE visualization of two phonemes /AA/ and /P/ with respect to the 7 accents, for layers 1, 5, 9, 12. $\theta_{ASR-LS}$, which shows similar patterns as $\theta_{ASR-L2}$, is excluded for brevity. See our web page for more phoneme examples.}
    \vspace{-4mm}
    \label{fig:vis}
\end{figure*}

\subsection{Implementation details}

For AID fine-tuning, we freeze the CNN layers in Wav2vec 2.0 and only update the Transformer layers, using an Adam optimizer with a fixed learning rate 1e-5. Other parameters are updated by a learning rate of 3e-3. The learning rate for regression heads is set to 1e-3. For all experiments, we train for 50 epochs, with a batch size of 48 utterances. 
Our implementation is based on the SpeechBrain toolkit \cite{ravanelli2021speechbrain}. 

\section{Results}
\subsection{Accent identification performance}
\label{res:aid}
Figure \ref{fig:aid} shows the accent-wise identification accuracy breakdown. In addition to fine-tuning, we also include an AID model that is trained with a weighted sum of the layer-wise SSL features, which follows the protocol of the SUPERB benchmark \cite{yang21c_interspeech} for downstream task evaluation. In this case, we freeze Wav2vec 2.0 and only update the layer-wise weights and the additional linear layers for AID. Fine-tuning provides strong performance on 5 out of the 7 accents. It outperforms the weighted sum approach in overall accuracy, while being worse on AB and SP accents. Due to the limited number of speakers in the training data, this may be caused by over-fitting. The non-trivial performance of the weighted sum approach indicates that the pre-trained Wav2vec 2.0 encodes useful features (e.g. phonetic and prosodic knowledge) for AID. 
\subsection{Phoneme probing}
We conduct probing experiments on 1) the pre-trained Wav2vec 2.0 ($\theta_0$); 2) the fine-tuned AID model ($\theta_{AID}$); 3) models that are fine-tuned by ASR tasks. For the third case, we consider 2 different ASR data, LibriSpeech \cite{libri} and L2-ARCTIC plus ARCTIC (same as AID training). These two models are denoted as $\theta_{ASR-LS}$ and $\theta_{ASR-L2}$, respectively. Since ASR tasks are linguistic-intensive, $\theta_{ASR-LS}$ and $\theta_{ASR-L2}$ serve as the references to quantify the linguistic knowledge learned in the AID task. On our AID test set, Word Error Rate (WER) of $\theta_{ASR-LS}$ and $\theta_{ASR-L2}$ is 23.99\% and 3.94\%, respectively.
%\footnote{The training and the test set share the \textit{same} prompts read by different speakers. Hence, this is a biased ASR training setting for $\theta_{ASR-L2}$.}
For all models, we probe the 12 Transformer layers.

\noindent\textbf{Does AID fine-tuning enable phonetic knowledge learning?} Figure \ref{fig:cca} shows the layer-wise CCA scores on utterances of 4 accents. The first 8 layers obtain similar CCA scores across all models. $\theta_{AID}$ has lower or comparable CCA scores than $\theta_0$ in layers 9 and 10, and significantly higher CCA scores in layers 11 and 12. The increase is even more prominent for $\theta_{ASR-LS}$ and $\theta_{ASR-L2}$ in layers 9-12. Higher CCA scores indicate more informative phoneme representations and a better ability to discriminate phonemes. This suggests that AID facilitates phonetic knowledge learning in the top 2 layers, with a similar effect as ASR fine-tuning. This finding provides evidence of the benefits of jointly modeling AID and ASR \cite{weninger2019deep, zhang2021accent}.

In addition, we observe that in layers 1-10, the gaps between $\theta_0$ and $\theta_{AID}$ are narrow on the US accent, while larger on other non-native accents (especially layers 8-10). Since $\theta_0$ is pre-trained on LibriSpeech which mainly contains native English speech, it is expected that phoneme representations learned on US-accented speech should be close to that learned on LibriSpeech. The fact that $\theta_{AID}$ representations on US accent tightly follow $\theta_0$ confirmed this point.

\noindent\textbf{Does the AID model learn accent-specific phoneme representations?}  Note that CCA scores are computed using canonical phoneme labels (i.e. the phonemes that are \textit{supposed to be pronounced}). On the other hand, L2 speakers usually vary in the pronunciations of specific phonemes. This motivates us to ask whether the gaps between $\theta_0$ and $\theta_{AID}$ on non-native accents suggest accent-specific phoneme representations. For this purpose, we use t-SNE \cite{vandermaaten08a} to visualize phoneme representations with respect to different accents. Figure \ref{fig:vis} presents the representations of vowel /AA/ and stop /P/ for layers 1, 5, 9, and 12.\footnote{We provide a web page for additional phoneme examples: \url{https://is23-2254.github.io/}} In general, we observe a gradual increase of accent-specificity in $\theta_{AID}$ starting from layer 1, and the strongest accent-specificity in layer 9.  Interestingly, in layer 12, $\theta_0$ also demonstrates some accent-specificity, especially in a few vowels such as /AA/, /AE/, while $\theta_{AID}$ shows stronger accent-specificity for stops such as /P/, /B/, fricatives such as /DH/, /SH/, affricates such as /CH/, /JH/ and diphthongs such as /OY/, /AY/. $\theta_{ASR-L2}$ looks accent-invariant, which is understandable because the ASR objective explicitly maps differently-accented phonemes to the same target. Corroborating with Figure \ref{fig:cca}, we conclude that $\theta_{AID}$ achieves both strong accent-specificity (as shown by the t-SNE visualization) and phoneme discrimination (as shown by the CCA scores) in layer 9. 

\begin{figure}[b!]
    \centering
    \vspace{-3mm}
    \includegraphics[width=1.0\columnwidth]{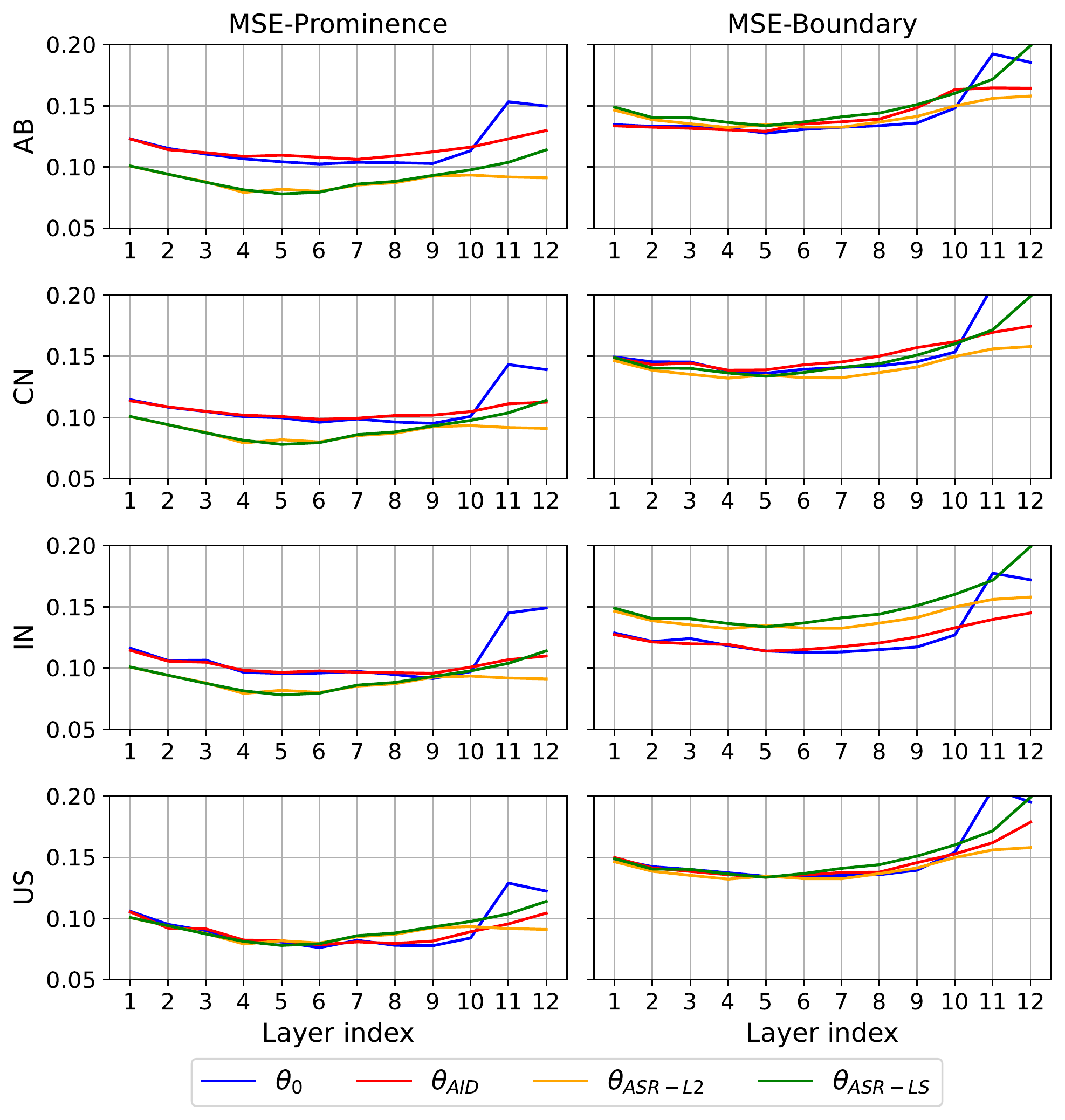}
    \vspace{-5mm}
    \caption{MSE of prominence and boundary prediction, for AB, CN, IN, US accents.}
    % \vspace{-5mm}
    \label{fig:mse}
\end{figure}

\subsection{Prosody probing}
\textbf{Does AID fine-tuning enable prosody knowledge learning?} Figure \ref{fig:mse} shows the MSE on prominence and boundary prediction. Comparing $\theta_{AID}$ to $\theta_0$, we find that both MSE scores differ mainly in top layers 11 and 12, while minimally in layers 1-10. This indicates that prosody knowledge is enhanced in the top 2 layers. Interestingly, we observe that in general ASR models achieve lower (better) MSE for prominence prediction in layers 1-8 (except for the US accent). One possible reason may be that prosodic prominence is closely related to the syntactic structure and lexical content of an utterance \cite{4358088}, which are also important for ASR tasks. Hence, when fine-tuning an ASR task on unseen accents, this knowledge may be enhanced implicitly as well for these accents. Nevertheless, in the case of the US accent, no further improvement is gained by ASR fine-tuning, which may suggest that the pre-trained model already learned such knowledge for the US accent during pre-training. This finding also sheds some light on the distribution of the prosodic prominence information in the 12 layers: given that the gaps primarily appear in layers 1-8, the prosodic prominence knowledge is likely to be encoded in these layers as well.

\section{Conclusions}
We have presented our study that aims to understand and quantify the phonetic and prosodic knowledge in a fine-tuned AID model. Through our phoneme and prosody probing analyses, we have shown that the AID fine-tuning task mainly causes changes in the top 4 layers. Layers 11 and 12 gained richer phoneme and prosody representation, while layer 9 showed strong accent-specific phoneme representations. Our study provides insights into the distribution of phonetic and prosody knowledge encoded in the Wav2vec 2.0 model, and how would an AID task affect such knowledge. Future works may include analyses of other SSL models with different architectures and training objectives.

\noindent\textbf{Acknowledgements} This study is supported by NSF EAGER CISE Project 2140415, and partially by the University of Texas at Dallas from the Distinguished University Chair in Telecommunications Engineering held by J. H. L. Hansen.

% \pagebreak
\bibliographystyle{IEEEtran}
\bibliography{mybib}

% Generated by IEEEtran.bst, version: 1.13 (2008/09/30)
\begin{thebibliography}{10}
\providecommand{\url}[1]{#1}
\csname url@samestyle\endcsname
\providecommand{\newblock}{\relax}
\providecommand{\bibinfo}[2]{#2}
\providecommand{\BIBentrySTDinterwordspacing}{\spaceskip=0pt\relax}
\providecommand{\BIBentryALTinterwordstretchfactor}{4}
\providecommand{\BIBentryALTinterwordspacing}{\spaceskip=\fontdimen2\font plus
\BIBentryALTinterwordstretchfactor\fontdimen3\font minus
  \fontdimen4\font\relax}
\providecommand{\BIBforeignlanguage}[2]{{%
\expandafter\ifx\csname l@#1\endcsname\relax
\typeout{** WARNING: IEEEtran.bst: No hyphenation pattern has been}%
\typeout{** loaded for the language `#1'. Using the pattern for}%
\typeout{** the default language instead.}%
\else
\language=\csname l@#1\endcsname
\fi
#2}}
\providecommand{\BIBdecl}{\relax}
\BIBdecl

\bibitem{weninger2019deep}
F.~Weninger, Y.~Sun, J.~Park \emph{et~al.}, ``Deep learning based mandarin
  accent identification for accent robust asr.'' in \emph{Proc. Interspeech},
  2019.

\bibitem{gao2021end}
Q.~Gao, H.~Wu, Y.~Sun \emph{et~al.}, ``An end-to-end speech accent recognition
  method based on hybrid ctc/attention transformer asr,'' in \emph{Proc.
  ICASSP}, 2021.

\bibitem{shi2021accented}
X.~Shi, F.~Yu, Y.~Lu \emph{et~al.}, ``The accented english speech recognition
  challenge 2020: open datasets, tracks, baselines, results and methods,'' in
  \emph{Proc. ICASSP}, 2021.

\bibitem{biadsy2011automatic}
F.~Biadsy, ``Automatic dialect and accent recognition and its application to
  speech recognition,'' Ph.D. dissertation, COLUMBIA UNIVERSITY, 2011.

\bibitem{witt2012automatic}
S.~M. Witt, ``Automatic error detection in pronunciation training: Where we are
  and where we need to go,'' in \emph{International Symposium on automatic
  detection on errors in pronunciation training}, 2012.

\bibitem{saito2022automated}
K.~Saito, K.~Macmillan, M.~Kachlicka \emph{et~al.}, ``Automated assessment of
  second language comprehensibility: Review, training, validation, and
  generalization studies,'' \emph{Studies in Second Language Acquisition},
  2022.

\bibitem{yang22v_interspeech}
M.~Yang, K.~Hirschi, S.~D. Looney \emph{et~al.}, ``{Improving Mispronunciation
  Detection with Wav2vec2-based Momentum Pseudo-Labeling for Accentedness and
  Intelligibility Assessment},'' in \emph{Proc. Interspeech}, 2022.

\bibitem{wu2021transformer}
M.~Wu, K.~Li, W.-K. Leung \emph{et~al.}, ``Transformer based end-to-end
  mispronunciation detection and diagnosis.'' in \emph{Proc. Interspeech},
  2021.

\bibitem{suzukida2021segmental}
Y.~Suzukida and K.~Saito, ``Which segmental features matter for successful l2
  comprehensibility? revisiting and generalizing the pedagogical value of the
  functional load principle,'' \emph{Language Teaching Research}, 2021.

\bibitem{kang2010suprasegmental}
O.~Kang, D.~Rubin, and L.~Pickering, ``Suprasegmental measures of accentedness
  and judgments of language learner proficiency in oral english,'' \emph{The
  Modern Language Journal}, 2010.

\bibitem{ulbrich2016prosody}
C.~Ulbrich and I.~Mennen, ``When prosody kicks in: The intricate interplay
  between segments and prosody in perceptions of foreign accent,''
  \emph{International Journal of Bilingualism}, 2016.

\bibitem{li2018automatic}
K.~Li, S.~Mao, X.~Li \emph{et~al.}, ``Automatic lexical stress and pitch accent
  detection for l2 english speech using multi-distribution deep neural
  networks,'' \emph{Speech Communication}, 2018.

\bibitem{deng21b_interspeech}
K.~Deng, S.~Cao, and L.~Ma, ``{Improving Accent Identification and Accented
  Speech Recognition Under a Framework of Self-Supervised Learning},'' in
  \emph{Proc. Interspeech}, 2021.

\bibitem{wang2021end}
D.~Wang, S.~Ye, X.~Hu \emph{et~al.}, ``An end-to-end dialect identification
  system with transfer learning from a multilingual automatic speech
  recognition model.'' in \emph{Proc. Interspeech}, 2021.

\bibitem{shon18_odyssey}
S.~Shon, A.~Ali, and J.~Glass, ``{Convolutional Neural Network and Language
  Embeddings for End-to-End Dialect Recognition },'' in \emph{Proc. Odyssey},
  2018.

\bibitem{baevski2020wav2vec}
A.~Baevski, Y.~Zhou, A.~Mohamed \emph{et~al.}, ``wav2vec 2.0: A framework for
  self-supervised learning of speech representations,'' \emph{Proc. NeurIPS},
  2020.

\bibitem{pasad2021layerwise}
A.~Pasad, J.-C. Chou, and K.~Livescu, ``Layer-wise analysis of a
  self-supervised speech representation model,'' in \emph{Proc. ASRU}, 2021.

\bibitem{dieck22_interspeech}
T.~tom Dieck, P.~A. Pérez-Toro, T.~Arias \emph{et~al.}, ``{Wav2vec behind the
  Scenes: How end2end Models learn Phonetics},'' in \emph{Proc. Interspeech},
  2022.

\bibitem{ma2021probing}
D.~Ma, N.~Ryant, and M.~Liberman, ``Probing acoustic representations for
  phonetic properties,'' in \emph{Proc. ICASSP}, 2021.

\bibitem{cormac-english-etal-2022-domain}
P.~Cormac~English, J.~D. Kelleher, and J.~Carson-Berndsen, ``Domain-informed
  probing of wav2vec 2.0 embeddings for phonetic features,'' in \emph{Proc.
  SIGMORPHON Workshop on Computational Research in Phonetics, Phonology, and
  Morphology}, 2022.

\bibitem{lin2022on}
G.-T. Lin, C.-L. Feng, W.-P. Huang \emph{et~al.}, ``On the utility of
  self-supervised models for prosody-related tasks,'' in \emph{Proc. SLT},
  2023.

\bibitem{feng2022silence}
C.-L. Feng, P.-c. Hsu, and H.-y. Lee, ``Silence is sweeter than speech:
  Self-supervised model using silence to store speaker information,''
  \emph{arXiv preprint arXiv:2205.03759}, 2022.

\bibitem{fan21_interspeech}
Z.~Fan, M.~Li, S.~Zhou \emph{et~al.}, ``{Exploring wav2vec 2.0 on Speaker
  Verification and Language Identification},'' in \emph{Proc. Interspeech},
  2021.

\bibitem{vaswani2017attention}
A.~Vaswani, N.~Shazeer, N.~Parmar \emph{et~al.}, ``Attention is all you need,''
  \emph{Proc. NeurIPS}, 2017.

\bibitem{suni2017hierarchical}
A.~Suni, J.~{\v{S}}imko, D.~Aalto \emph{et~al.}, ``Hierarchical representation
  and estimation of prosody using continuous wavelet transform,''
  \emph{Computer Speech \& Language}, 2017.

\bibitem{belinkov2017analyzing}
Y.~Belinkov and J.~Glass, ``Analyzing hidden representations in end-to-end
  automatic speech recognition systems,'' \emph{Proc. NeurIPS}, 2017.

\bibitem{amodei2016deep}
D.~Amodei, S.~Ananthanarayanan, R.~Anubhai \emph{et~al.}, ``Deep speech 2:
  End-to-end speech recognition in english and mandarin,'' in \emph{Proc.
  ICML}, 2016.

\bibitem{prasad2020accents}
A.~Prasad and P.~Jyothi, ``How accents confound: Probing for accent information
  in end-to-end speech recognition systems,'' in \emph{Proc. ACL}, 2020.

\bibitem{triantafyllopoulos22b_interspeech}
A.~Triantafyllopoulos, J.~Wagner, H.~Wierstorf \emph{et~al.}, ``{Probing speech
  emotion recognition transformers for linguistic knowledge},'' in \emph{Proc.
  Interspeech}, 2022.

\bibitem{vaidya2022deep}
M.~Vaidya, K.~Sabu, and P.~Rao, ``Deep learning for prominence detection in
  children’s read speech,'' in \emph{Proc. ICASSP}, 2022.

\bibitem{kominek2004cmu}
J.~Kominek and A.~W. Black, ``The cmu arctic speech databases,'' in \emph{Fifth
  ISCA workshop on speech synthesis}, 2004.

\bibitem{zhao2018l2}
G.~Zhao, S.~Sonsaat, A.~Silpachai \emph{et~al.}, ``L2-arctic: A non-native
  english speech corpus.'' in \emph{Proc. Interspeech}, 2018.

\bibitem{szegedy2016rethinking}
C.~Szegedy, V.~Vanhoucke, S.~Ioffe \emph{et~al.}, ``Rethinking the inception
  architecture for computer vision,'' in \emph{Proc. CVPR}, 2016.

\bibitem{kakouros2022speech}
S.~Kakouros, T.~Stafylakis, L.~Mosner \emph{et~al.}, ``Speech-based emotion
  recognition with self-supervised models using attentive channel-wise
  correlations and label smoothing,'' \emph{arXiv preprint arXiv:2211.01756},
  2022.

\bibitem{choi2021neural}
H.-S. Choi, J.~Lee, W.~Kim \emph{et~al.}, ``Neural analysis and synthesis:
  Reconstructing speech from self-supervised representations,'' \emph{Proc.
  NeurIPS}, 2021.

\bibitem{qian2022contentvec}
K.~Qian, Y.~Zhang, H.~Gao \emph{et~al.}, ``Contentvec: An improved
  self-supervised speech representation by disentangling speakers,'' in
  \emph{Proc. ICML}, 2022.

\bibitem{mcauliffe17_interspeech}
M.~McAuliffe, M.~Socolof, S.~Mihuc \emph{et~al.}, ``{Montreal Forced Aligner:
  Trainable Text-Speech Alignment Using Kaldi},'' in \emph{Proc. Interspeech},
  2017.

\bibitem{ravanelli2021speechbrain}
M.~Ravanelli, T.~Parcollet, P.~Plantinga \emph{et~al.}, ``Speechbrain: A
  general-purpose speech toolkit,'' \emph{arXiv preprint arXiv:2106.04624},
  2021.

\bibitem{yang21c_interspeech}
S.-W. Yang, P.-H. Chi, Y.-S. Chuang \emph{et~al.}, ``{SUPERB: Speech Processing
  Universal PERformance Benchmark},'' in \emph{Proc. Interspeech}, 2021.

\bibitem{libri}
V.~Panayotov, G.~Chen, D.~Povey \emph{et~al.}, ``Librispeech: An asr corpus
  based on public domain audio books,'' in \emph{Proc. ICASSP}, 2015.

\bibitem{zhang2021accent}
Z.~Zhang, Y.~Wang, and J.~Yang, ``Accent recognition with hybrid phonetic
  features,'' \emph{Sensors}, 2021.

\bibitem{vandermaaten08a}
L.~van~der Maaten and G.~Hinton, ``Visualizing data using t-sne,''
  \emph{Journal of Machine Learning Research}, 2008.

\bibitem{4358088}
S.~Ananthakrishnan and S.~S. Narayanan, ``Automatic prosodic event detection
  using acoustic, lexical, and syntactic evidence,'' \emph{IEEE Transactions on
  Audio, Speech, and Language Processing}, 2008.

\end{thebibliography}

\end{document}